# INTRODUCING COMPUTATIONAL THINKING IN CALCULUS FOR ENGINEERING

J. Bilbao    E. Bravo    O. Garcia    C. Rebollar

*Applied Mathematics Department, University of the Basque Country (UPV/EHU), Bilbao, Spain*
*javier.bilbao@ehu.eus, eugenio.bravo@ehu.eus, olatz.garcia@ehu.eus, carolina.rebollar@ehu.eus*

**Abstract-** Technology is currently ubiquitous and is also part of the educational system at all levels. It started with communication technology systems, and later continued with digital competence. Nowadays, although these previous concepts are still in force and are useful for students and workers in general, a new concept has been born that can function as a cross-curricular competence called Computational Thinking. There is currently no consensus on the definition of computational thinking, nor on the classification of its skills, but there is a consensus that it refers to a set of skills necessary for the formulation and resolution of problems. The study of Computational Thinking has been very influential in recent years in research on teaching and learning processes, which has led educational institutions to begin to address these issues during training. In this paper, we try to introduce this new cross-curricular competence and expose a project of implementation of Computational Thinking in engineering careers through Calculus subject.

**Keywords:** Computational Thinking, Education, Engineering, Competences, Curriculum.

## 1. INTRODUCTION

The irruption of computational thinking in the educational system is becoming more and more notorious, with its inclusion in the curricula of several countries [1, 2], and as a current topic in research [3-5]. In recent decades, society has been transformed at all levels: social, economic, cultural and also educational. The influence initially of Information and Communication Technologies (ICT), and later of digital competence, in this transformation is undeniable, as well as the uncertainty that all this creates when talking about the future [6]. On the other hand, in recent years there has been an increased interest in the teaching of computer science in pre-university educational stages, both for purely educational and economic reasons [7, 8]. Moreover, in different educational settings across countries, a consensus is emerging on the importance of computer science education, and that it should not be limited to the development of digital competence. Thus, students should also acquire basic knowledge of computer science as a science and technology [9, 10].

National responses to the requirements for students to acquire digital and computing competencies are very varied [11-13], ranging from compulsory computer science subjects in national or regional curricula to informal or non-formal initiatives to their cross-curricular teaching. At the university stage, the learning of the contents of Computer Science as a basic science in Engineering is usually focused on contributing to develop in students of Engineering careers in general the ability to build the solution of problems with the computer (through the ingenuity of algorithms and their codification as programs) [14]. The greatest cognitive challenge faced by students is the discovery of an effective and efficient algorithm that solves the problem posed [15].

The objectives of this type of subjects related to Computer Science can generally be grouped into four main blocks. In the first, the student is expected to understand the fundamentals of current technology and how it influences computer design. In the second, the student should gain an understanding of how system software controls the operation of a computer and establishes the basic communication paths between a machine and the people who use it, and how computers can be connected to share information and resources. The third block is concerned with training the student to discover and represent algorithms as programs, applying the divide-and-conquer principle and successive refinements, and instilling in him the search for efficiency and checking for correctness and completeness. The fourth block is usually devoted to the programming process, to the communication of algorithms to computers, currently using languages such as Python, studying also the design of data to represent information and manipulate it with the greatest efficiency, and the effective presentation of information to the user. The year 2006 saw the emergence of a new concept, computational thinking, which over the years has made its way as a new cross-curricular competence in the educational system of most countries in the world.

## 2. COMPUTATIONAL THINKING

One of the fields in which work is being done for the introduction of computational thinking in the curricula is its link with Mathematics. Although computational thinking is a transversal competence that can be used in any field (STEAM), it is no less true that it may seem more related to the more technological subjects (STEM). Thus, some steps have been taken to relate this new competence (CT) to some subject or area in which its relationship can be seen more clearly. For example, in Spain, CT has begun to be introduced in mathematics subjects in primary and secondary education [16]. Or in the 2022 PISA tests, it was related to mathematical reasoning [17]. There are also research projects to analyze the relationship of CT with algebraic thinking [18]. Computational Thinking is the set of Thought Processes that develop and formulate the situations and problems that may arise. This development includes the representation of these possible solutions. In addition, the whole process can be implemented so that an information processing agent (human, computer or combinations of human and computer) can carry out all the steps and obtain the solution(s).

Computational Thinking, abbreviated CT, like, for example, arithmetic, is a Skill and an Attitude of Universal application for all people. People use CT when reading, writing, speaking and listening, when studying mathematics, history, etc., as well as in their personal and professional lives. As a transversal competence, it has several concepts in common with some methodologies, such as ABP (Problem Based Learning), and is closely related to some fields, such as STEAM. Concepts such as "knowledge construction" are included in the DNA of Computational Thinking, but that does not imply that they are its property. That is, there are clear relationships between Computational Thinking and some basic methodologies.

According to OECD [19], education systems and their curricula should help students, and anyone else in their lifelong learning process, to be, on the one hand, responsible users of technology, and, on the other hand, although less common for most people, to be able to generate technology for their own or society's use. To do so, they must develop the necessary skills and competencies and be in line with the new forms of today's economy and society. It should be remembered that, empowered by technology, knowledge management is an important competency for the management of knowledge. In the aforementioned report, these skills were defined to include processes related to the networked environments. Furthermore, it was stated that these skills should be started to teach and learn at school. The definition does not focus on the subjects in which digital competence has to be taught or the subjects it affects. Therefore, it is a field that can be interpreted as generalist, including, in addition to the subjects, the educational levels where it should be taught.

Computational Thinking is a kind of analytical thinking in which both mathematical thinking and engineering concepts are used in order to understand any problem that may be encountered in different areas of knowledge and, subsequently, to be able to solve it. This concept of CT was used at the end of the 20th century by S. Papert [20]. Papert focused more on the area of Computer Science, where for example he developed the Logo software. However, in 2006, Wing [21] introduced Computational Thinking into society, through Computer Science, to push that field forward and, according to her, to describe how a computer scientist should think. In her article, Wing defined Computational Thinking as problem solving, system design and understanding human behavior using the fundamental concepts of computer science.

## 3. NEW COMPETENCE

Computational thinking has entered strongly as part of the curriculum, especially, but not only, within the areas of knowledge that have to do with the so-called STEM. For example, within Engineering, general competencies can be summarized as the knowledge of basic technologies and methods that enable them to learn new methods and technologies, as well as providing them with the versatility to adapt to new situations [22]; and the ability to solve problems with initiative, decision making, creativity, and to communicate and transmit knowledge, skills and abilities, understanding the ethical and professional responsibility of the engineer's activity.

The main skills of Computational Thinking are currently considered to include: abstraction, decomposition, data collection, analysis and representation, algorithmic thinking, transferability, and evaluation and adjustment.

In the case of algorithm thinking, this form of thinking involves the ability to decompose a complex problem into a series of simple steps or instructions, organized in a logical and orderly manner, to solve it effectively and efficiently. In other words, algorithmic thinking is the ability to design and create algorithms [23].

An algorithm is a set of precise and ordered instructions used to perform a task or solve a specific problem. These instructions are usually sequential and must follow a particular order, and may also include conditions, loops and other programming elements.

Algorithmic thinking is applied in many areas of life, not just programming. For example, when planning a party, algorithmic thinking can be used to break down tasks into simple, organized steps, such as sending invitations, buying food and drink, decorating the space, and preparing the music. In this way, you can ensure that all aspects of the party are covered and are carried out effectively and efficiently.

In the context of programming, algorithmic thinking is a fundamental skill for designing and creating software. A programmer must be able to break down a problem into logical and orderly steps, design an effective and efficient algorithm to solve it, and then implement that algorithm in a specific programming language.

In short, algorithmic thinking is a fundamental skill in programming and in many other aspects of life. It is the ability to decompose a complex problem into simple and ordered steps, design an effective and efficient algorithm to solve it, and then implement that algorithm in a specific language.

Within the Algorithmic Thinking skill, it is interesting to define these two concepts:
- Modeling:
This is about establishing the steps to be followed to solve the problem until the solution is reached. Modeling is often done by creating mathematical models or simulations that allow exploring different scenarios and evaluating possible solutions.
- Automation:
This is about doing repetitive or tedious tasks with or like a computer, in order to save labor and time, thus being more efficient.

Both problem modeling and task automation are essential components of Algorithmic Thinking, as they help decisively in the design of the problem solving process, allowing repetitive and tedious tasks to be carried out effectively, thus saving time and reducing errors.

## 4. CLASSROOM EXPERIENCE

Curriculum change in higher education is usually laborious, controversial and always exposed to different opinions and sensitivities. We do not want to excuse or criticize these processes in this paper, but to state that, being difficult the introduction of a new subject, such as computational thinking, there are other ways to teach concepts, such as joining. Thus, in the Calculus course of the Engineering degree in Industrial Technology Engineering, degree in Industrial Organization Engineering and degree in Environmental Engineering, computational thinking has been introduced to prepare students to solve some problems of the subject. We have conducted several tests in class with engineering students, using problems, tests, and also measuring the capacity of computational thinking from exercises of engineering subjects, such as Calculus. The objective is to improve the teaching and learning processes of the competencies involved in Computational Thinking, so that future engineers are better prepared to take on the challenges demanded by contemporary society. The viability of the project is based on the advances obtained through previous works of the members in the area and their interaction with other European research groups.

It is proposed to develop didactics for the teaching of computational thinking for engineering students and at the same time to propose active methodologies for the use and delivery of materials in face-to-face and online educational experiences. A few years ago, the first experiences began in the subject Calculus, which is taught in the first year of Engineering degrees. Material was developed for these subjects focused on the main knowledge of Computational Thinking, such as abstraction, decomposition, algorithmic thinking, etc. In the practices, sometimes the strategy of learning based on problems is used, as part of the improvement with active methodologies. The aim is to motivate students in different careers by developing problems that are useful to them in their profession.

The idea is that students understand, through the development of these problems, the importance of these skills for their future professional and personal life. The problems are subsequently evaluated by the teacher, who, taking into account the training of the students, since they are carried out before and after the possible use of the CT. With this innovation plan, we seek that our graduates can make use of technologies to solve the problems they face in their professional activity, to provide solutions to an increasingly demanding society, which demands faster and more efficient answers. The term "computational thinking" was first introduced in 2006 by Jeannette M. Wing in a short article entitled Computational Thinking. She conceives of it as a discipline that involves "solving problems, designing systems, and understanding human behavior, using concepts that are fundamental to computer science." In short, it is a philosophy of posing and solving problems using the logic by which machines are governed.

Wing herself expanded her definition in 2011 along these lines: "Computational thinking is the mental processes involved in formulating problems and their solutions so that the solutions are represented in such a way that they can be carried out effectively by an information processing agent".

Two concepts emerge from this definition: that it is a form of reasoning that does not depend exclusively on technology, and that it is a methodology for problem solving by humans, by machines, or through the collaboration of both. Basically, it consists of posing a problem following the operational process of an intelligent system. Computational thinking is not programming, or even thinking like a computer does; it is a way of solving problems. Our effort is focused on the introduction of computational thinking in formal education not being limited to computer science and the more technical areas, but being a transversal education, applicable to different fields of knowledge.

## 5. METHODOLOGY

The evaluative model acquires, among others, a series of specifications regarding the delimitation of objectives, hypotheses, data collection techniques and analysis of the data obtained. These specifications in our work are always focused on the two fundamental contents to be evaluated: analyzing the process of introducing computational thinking in the Calculus subject and the results obtained by the students during the implementation of the process.

The study was carried out in three groups of students. Each group had a different teacher. The number of students in each group was not homogeneous, since one group, G1, had 56 students, another, G2, with 76 students, and the third, G3, with 49. Therefore, there were a total of 181 students. Table 1 summarizes the final distribution.

Table 1. Distribution of students

| G1 | G2 | G3 |
|----|----|----|
| 56 | 76 | 49 |

The evaluation of the process of introducing computational thinking in the subject of Calculus has as its fundamental objective to know the degree of assimilation or incorporation of said procedure in the students. Computational thinking has several skills that, usually, do not need to be applied all at once to solve a problem. Therefore, in this study, we focus on the algorithmic thinking part, and with more specificity on its modeling part.

Thus, this general objective will be specified in the following more specific objectives:
a) Know if the student differentiates the problem-solving strategy: understanding the problem and executing a problem modeling strategy.
b) Know the integration that the student has made of computational thinking that helps him differentiate the three parts of the problem-solving process (what we know about this problem, what we want to know, what needs to be put together, removed or distributed, etc.).
c) Know what degree of assimilation of computational thinking has been acquired through its application to understand and solve the problem.
d) Know the difference in the application of modeling within algorithmic thinking.
e) Know if there are differences between the subjects in understanding, execution, verification and degree of introduction of computational thinking according to the teacher who has implemented the process.
g) Know if there are differences between the subjects in terms of understanding, execution and verification of the introduction of computational thinking according to the level of resolution (high, medium and low) of the subjects.

In order to achieve the stated objectives, data collection was proposed based on the observation of the introduction of computational thinking through exercises from the Calculus subject. In this paper, one of these exercises is presented.

We present below a Calculus exercise in which the mathematical concepts of continuous function, derivative, differential, and directional derivative come into play. However, what we are going to analyze is the part of computational thinking related to the skill of algorithmic thinking, with a specific section on modeling. With this, we want to show that computational thinking can also be evaluated in exercises that are not ad-hoc for the competence.

Given the following function:

$$f(x,y) = \begin{cases} \sin\left(\dfrac{xy^2}{x^2+y^2}\right) & \forall (x,y) \neq (0,0) \\ 0 & (x,y) = (0,0) \end{cases} \quad (1)$$

a) Study the continuity of the function at the origin.
b) Calculate the partial derivatives of the function at the origin.
c) Study the differentiability of the function at the origin.
d) Calculate the directional derivative of the function at the origin according to the direction given by the vector.

## 6. RESULTS AND DISCUSSION

A typical problem from the Calculus subject for engineering related to the part of continuity and differentiability was used, in which the study of a certain function at a point was requested. In addition to the evaluation of the mathematical component, the computational thinking present in said problem was evaluated. Since computational thinking is made up of several different skills, and studying all of them at once would be an excessive length for a single publication, we have focused on one of them: algorithmic thinking. Furthermore, within algorithmic thinking, modeling can be defined. As stated above, this is about establishing the steps to be followed to solve the problem until the solution is reached. Modeling is often done by creating mathematical models or simulations that allow exploring different scenarios and evaluating possible solutions.

In our study, and for the problem that we pose to the students, we have carried out the evaluation and measured results based on the following classification.

On the one part, the modeling of the problem, that is, its formulation, consisted of: study continuity, partial derivatives, differentiation and directional derivative at the point for the explicit function. Its stages are the following:
1) Student studies continuity at the point. To analyze continuity, it is necessary that the limit at the point coincides with the value of the function at the point. This item has been called M1.1.
2) Student studies the partial derivatives at the point. The partial derivatives are necessarily calculated through the limit that defines them. This item has been called M1.2.
3) Student studies differentiation at the point. If the function is not continuous or does not have any partial derivatives, then it is not differentiable. Otherwise, it will be necessary to check if it is differentiable through the necessary and sufficient condition. This item has been called M1.3.
4) Student calculates the directional derivative at the point. If the function is differentiable, the directional derivative is calculated as the dot product of the gradient vector and the normalized vector. Otherwise, the directional derivative is calculated through the definition. This item has been called M1.4.

On the other part, the algorithmic thinking skill was evaluated through the following items (at the end of each item the code that we have assigned appears):
1) Student solves each of the 4 stages in an orderly manner (PA1.1).
2) Student checks the conditions to be able to apply each criterion or theorem (PA1.2).
3) Student correctly applies each criterion or theorem (PA1.3).
4) Student correctly calculates the limits (PA1.4).
5) Student correctly calculates partial derivatives (PA1.5).
6) Student correctly calculates the directional derivative (PA1.6).
7) Student operates correctly with vectors (PA1.7).

Table 2. Evaluation rubric - The first row corresponds specifically to the modeling evaluation - The second row corresponds to the evaluation of algorithmic thinking

| Poor (0) | Insufficient (1) | Enough (2) | Advanced (3) | Excellent (4) |
|---|---|---|---|---|
| Student does not formulate the problem to be solved. Student does not establish the stages to follow to reach the solution. | Student formulates the problem to solve. Student does not establish the stages to follow to reach the solution. | Student formulates the problem to be solved. Student does not correctly establish all the stages to follow. | Student formulates the problem to be solved. Student correctly establishes all the stages to follow. | Student formulates the problem to be solved. Student establishes the best model with all the stages to follow. |
| Student does not establish the steps to follow to solve the problem. Student does not perform any repetitive tasks automatically | Student establishes, but does not order, the steps to follow to solve the problem. Student performs some of the repetitive tasks automatically | Student orders the steps to follow, although the solution to the problem is not reached. Student performs some of the repetitive tasks automatically | Student orders the steps to follow to solve the problem. Student performs most repetitive tasks automatically | Student orders in the best possible way the steps to follow to solve the problem. Student performs all repetitive tasks automatically |

All the items, both for modeling and for algorithm thinking, was evaluated according to the rubric shown in Table 2.

Figure 1 shows the arithmetic mean of the complete exam grade of each of the three groups (out of 10 points), while Figure 2 shows the mean of the problem analyzed in this study. Although the pattern followed by both averages is similar, with group 31 having better grades, the difference between the groups decreases.

Furthermore, it can be seen that the group means are slightly above half the value of the item, both when it corresponds to the complete exam and when it corresponds only to the analyzed exercise, which was evaluated out of 2.5 points. It would be logical to think, then, that the evaluation of computational thinking follows a similar pattern.

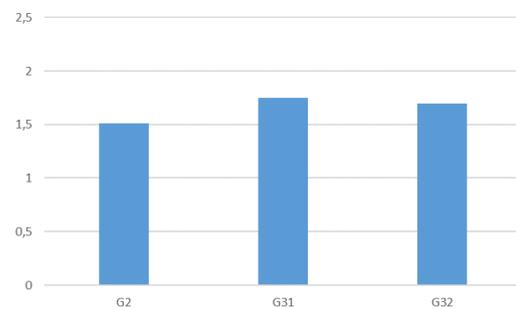

Figure 2. Arithmetic mean of the problem of Calculus analyzed for each of the three groups. The grade is out of 2.5 points

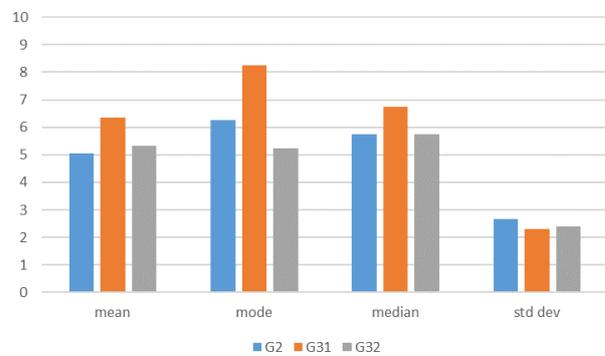

Figure 3. Arithmetic mean, mode, median and standard deviation for each of the groups, calculated with respect to the overall exam grade for the Calculus subject (out of 10 points)

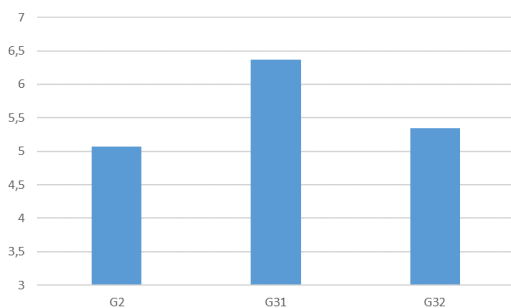

Figure 1. Arithmetic mean of the complete exam grade of Calculus of each of the three groups. Notice that, although the exam is out of 10 points, the scale of the graphic is from 3 to 7 for a better visualization

Figure 3 shows different statistics for each of the groups, calculated with respect to the overall exam grade for the Calculus subject. As can be seen, there are differences between the different groups, although they are not substantial. One of the aspects that can be highlighted is that the average grade is in the range between 5 and 6.3 points out of 10. The conceptual demand required in the subject is high and that is why the application of computational thinking as an improvement of the teaching-learning process can be a valuable tool.

As has been said, one of the skills that have been measured through exercises specific to the Calculus subject has been algorithmic thinking. Furthermore, an important part of that skill, such as modeling, has been evaluated as a separate item. Both evaluations have been measured on 4 points, as the rubric presented in Table 2.

Figure 4 shows the results of these evaluations for each of the student groups. It can be seen that, on the one hand, the modeling follows a trend very similar to that of the average grade for the Calculus exam. But on the other hand, the evaluation of algorithm thinking has been different, and this factor may be related to the students in each group and to the teacher or his methodology.

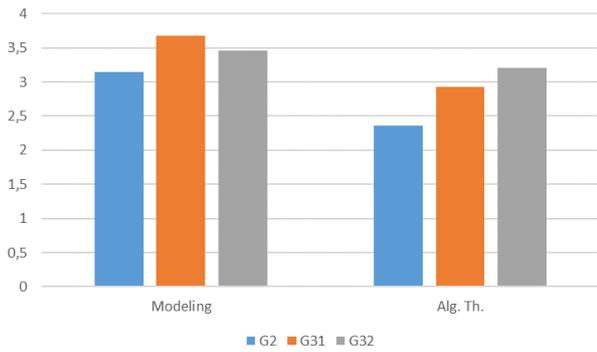

Figure 4. Evaluation of modeling and algorithmic thinking in each group (out of 4 points)

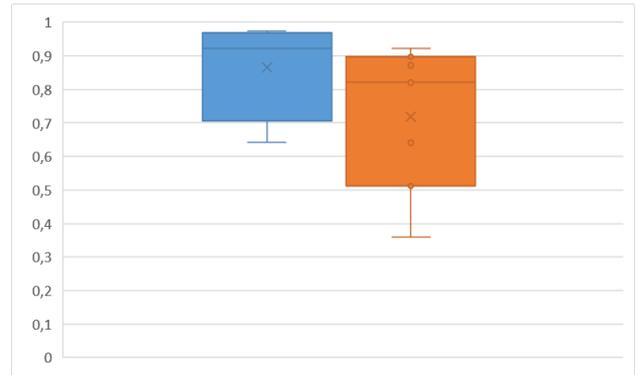

Figure 5. Boxplot, corresponding to group 32, of the modeling (blue) and algorithm thinking (red) items. Each item is worth 1 point

According to the previous comment, we can see in Table 3 that the correlation of the two grades, the exam grade and the exercise grade, with the modeling evaluation is very high, even higher than between the exam grade and the exercise grade.

However, the correlation of algorithmic thinking is, on the one hand, low (0.32) with the exam grade, while it is significant (0.69) with the exercise.

Obviously, not all the exam exercises required the same "amount" of algorithmic thinking as the one presented in this paper, although they did require the appropriate formulation, modeling, for their resolution. Furthermore, we must not forget that the average of the students' grades is slightly higher than half of what is possible.

Table 3. Correlation matrix (all groups)

|            | mean exam   | mean exerc | Model      | AlgTh |
|------------|-------------|------------|------------|-------|
| mean exam  | 1           |            |            |       |
| mean prob  | 0.80200588  | 1          |            |       |
| Model      | 0.90994388  | 0.9775061  | 1          |       |
| AlgTh      | 0.3297659   | 0.8283781  | 0.69160092 | 1     |

If we focus only on the modeling, and if we observe the items that make it up in the presented evaluation, we see disparate correlations, as it is shown in Table 4.

This values of correlation may indicate the lesser or greater importance of some of these items in achieving the correct result of the exam and the exercise, being very small in the case of M1.1, corresponding to the mathematical definition of continuity.

Table 4. Correlation matrix (G32 - Model)

|       | exam   | exerc  | M1.1   | M1.2   | M1.3   | M1.4   | Model |
|-------|--------|--------|--------|--------|--------|--------|-------|
| exam  | 1      |        |        |        |        |        |       |
| exerc | 0.7786 | 1      |        |        |        |        |       |
| M1.1  | 0.3011 | 0.3577 | 1      |        |        |        |       |
| M1.2  | 0.4444 | 0.5126 | 0.6977 | 1      |        |        |       |
| M1.3  | 0.629  | 0.6357 | 0.4799 | 0.6877 | 1      |        |       |
| M1.4  | 0.4841 | 0.592  | 0.2168 | 0.3107 | 0.4518 | 1      |       |
| Model | 0.6271 | 0.7171 | 0.6229 | 0.7638 | 0.8294 | 0.7982 | 1     |

The boxplot in Figure 5 further explains the difference in modeling and algorithmic thinking, the latter having a larger range of values and being, in general, inferior to modeling.

## 7. CONCLUSIONS

In recent times, much emphasis has been placed on the need for computer programming to become a curricular discipline from the most basic educational levels, in order to prepare students to be able to live, and above all, to be able to work, in a world in which technology is omnipresent. There is no doubt that learning programming languages can provide us with knowledge about the logical structure of a computer system's operation, but simply teaching programming can be limited and even insufficient. Thus, the introduction of computational thinking in educational systems, from the early years to the university stage, can improve students' preparation, not only in some subjects, but also in their preparation for professional life in any area of knowledge.

## ACKNOWLEDGEMENTS

The authors are grateful to Concepcion Varela for her collaboration in the development and implementation of the paper. The authors thank the support of the laboratory of the Applied Mathematics Department of the University of the Basque Country (UPV/EHU), Bilbao, Spain in the development of the study.

## BIOGRAPHIES

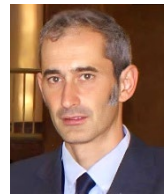

Name: **Javier**
Surname: **Bilbao**
Birthday: 1967
Birthplace: Spain
Master: Industrial Engineering, University of the Basque Country, Spain, 1991

Doctorate: Ph.D., Applied Mathematics, University of the Basque Country, Spain
The Last Scientific Position: Prof., Applied Mathematics Department, Engineering School of Bilbao, University of the Basque Country, Spain
Research Interests: Distribution Overhead Electrical Lines Compensation, Optimization of Series Capacitor Batteries in Electrical Lines, Modulization of a Leakage Flux Transformer, Losses in Electric Distribution Networks, Artificial Neural Networks, Modulization of Fishing Trawls, E-Learning, Noise of Electrical Wind Turbines, Light Pollution, Machine Learning, Computational Thinking
Scientific Publications: 76 Papers, 30 Books, 1 Patent, 48 Projects, 161 Conference Papers
Scientific Memberships: Bebras Community, IOTPE Organization

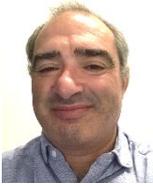

Name: **Eugenio**
Surname: **Bravo**
Birthday: 1967
Birthplace: Spain
Master: Industrial Engineering, University of the Basque Country, Spain, 1991
Doctorate: Ph.D., Electrical Engineering, University of the Basque Country, Spain
The Last Scientific Position: Prof., Applied Mathematics Department, Engineering School of Bilbao, University of the Basque Country, Spain
Research Interests: Distribution Overhead Electrical Lines Compensation, Optimization of Series Capacitor Batteries in Electrical Lines, Modulization of a Leakage Flux Transformer, Losses in Electric Distribution Networks, Artificial Neural Networks, Modulization of Fishing Trawls, E-Learning, Noise of Electrical Wind Turbines, Computational Thinking
Scientific Publications: 56 Papers, 25 Books, 1 Patent, 38 Projects, 135 Conference Papers
Scientific Memberships: Bebras Community, IOTPE Organization

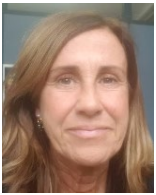

Name: **Olatz**
Surname: **Garcia**
Birthday: 1966
Birthplace: Spain
Master: Mathematics, University of the Basque Country, Spain
Doctorate: Science, Mathematics section, University of the Basque Country, Spain
The Last Scientific Position: Prof., Applied Mathematics Department, Engineering School of Bilbao, University of the Basque Country, Spain
Research Interests: E-Learning, Optimization of Series Capacitor Batteries in Electrical Lines, Noise of Electrical Wind Turbines, Machine Learning, Computational Thinking
Scientific Publications: 45 Papers, 25 Books, 1 Patent, 33 Projects, 111 Conference Papers
Scientific Memberships: Bebras Community, IOTPE Organization

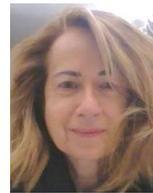

Name: **Carolina**
Surname: **Rebollar**
Birthday: 1963
Birthplace: Spain
Master: Mathematics, University of the Basque Country, Spain
Doctorate: Science, Mathematics section, University of the Basque Country, Spain
The Last Scientific Position: Prof., Applied Mathematics Department, Engineering School of Bilbao, University of the Basque Country, Spain
Research Interests: E-Learning, Noise of Electrical Wind Turbines, Machine Learning, Computational Thinking
Scientific Publications: 36 Papers, 24 Books, 1 Patent, 25 Projects, 87 Conference Papers
Scientific Memberships: Bebras Community, IOTPE Organization, SEMA